\documentclass[aps,prb,twocolumn,superscriptaddress]{revtex4-2}
\usepackage{graphicx}                % 图片处理（保留）
\usepackage{subcaption}              % 替代过时的subfigure包（关键修改）
\usepackage{enumitem}      \usepackage{makecell}    % 核心：表格多行单元格

\usepackage{float}                   % 浮动体控制（解除注释）
\usepackage[
    colorlinks,
    bookmarks=false,
    citecolor=blue,
    linkcolor=blue,
    urlcolor=blue,
    breaklinks=true                  % 新增：自动换行链接
]{hyperref}
\usepackage{braket}
\usepackage{bm}                      % 粗体数学符号（保留）
\usepackage{amsmath,amssymb,amsfonts,amsthm} % 数学公式/符号/定理（合并加载）
\usepackage{caption}

\usepackage{tikz}
\usepackage{pdfpages}
\makeatletter
\AtBeginDocument{\let\LS@rot\@undefined}
\makeatother
 \usepackage{ragged2e}
\setlist{nosep}       
               
\begin{document}

% 7. 标题（优化排版，支持换行）
\title{Localization Phase diagram of the Hexagonal Lattice with Irrational Magnetic Flux}

% 8. 作者信息（规范格式 + 新增注释） 
\title{Localization phase diagram of the Hexagonal Lattice with irrational magnetic flux}

\author{Qi Gao}
\affiliation{National Laboratory of Solid State Microstructures and School of Physics, Nanjing University, Nanjing, China}

\author{Shuo Zhang}
\affiliation{School of Mathematics, Nanjing University, Nanjing, China}

\author{Wei Chen}
\email{chenweiphy@nju.edu.cn}  % 通讯作者邮箱
\affiliation{National Laboratory of Solid State Microstructures and School of Physics, Nanjing University, Nanjing, China}
\affiliation{Collaborative Innovation Center of Advanced Microstructures, Nanjing University, Nanjing China}

\begin{abstract}
We study the Hofstadter model on a hexagonal lattice with irrational magnetic flux in this work. The Hofstadter model of the square lattice with irrational flux has been solved mathematically by Avila and his collaborators  in his Fields medal work. However, this theory is usually not applicable to lattices with internal degrees of freedom, such as spin or sublattices. In this work, we show that for the hexagonal lattice with only nearest neighbor hoppings, the system can still be characterized by a $2*2$ transfer matrix and solved exactly by Avila's global theory although this lattice has two sublattices. We obtained the exact localization phase diagram of the hexagonal lattice with irrational flux by this theory, which reveals three pure phases, i.e., the extended, localized and critical states but no mobility edge due to the chiral symmetry. We used the renormalization group (RG) theory to verify these results, which can determine part of the phase diagram. We then computed the fractal dimension of the remaining part numerically. The results from both the RG theory and numerical analysis confirmed the phase diagram we get from Avila's global theory. Our results can be tested in various hexagonal Moire lattices and artificial superlattices in recent experiments. 
\end{abstract}
 
\maketitle

In 1976, Hofstadter first revealed that the energy spectrum of electrons in a square-lattice under a uniform magnetic field with commensurate flux exhibits a striking self-similar fractal pattern, which is now widely known as the Hofstadter spectrum~\cite{Hofstadter_1976}. With the discovery of the quantum Hall effect \cite{Von-Klitching}, the Hofstadter model with rational flux was considered  as a fundamental model for this effect in a lattice and has been studied intensively in the past decades \cite{Hofstadter_k_Bernevig_2020,Hexagonal_Wangjian_2020,WEN1989,Xiao2024,Gao_Chen_2026,Square(NNN)_Kohmoto_1990,Zeromodes_Kohmoto_1989,Symmetry_indicators,Dimerized,Hofstadter_Real,Zak,Many_Chern}. In contrast, the Hofstadter model with irrational flux is much less investigated. However, the Hofstadter model with irrational flux is universal due to the fluctuation in a real system \cite{Thoulss_Bandwidths_1983,Thouless1990,thouless1991total,APT_Thouless_1994,S.J_PRL_1995,S.J_PRL_1996,AAH_1980}. The study of the irrational flux case is then of fundamental meaning. Moreover, the Hofsadter model with irrational flux itself also demonstrates interesting and rich physics and has attracted growing interest in recent years \cite{Unification,Jitomirskaya2012,SquareNNN_2017,Avila2015,Simon_2021,B.Hetenyi_1,B.Hetenyi_2}.

So far the study of the Hofstadter model with irrational flux is mainly on the square lattice. In 1994, Thouless {\it et al.} identified the phase diagram of electrons in the square-lattice Hofstadter model of irrational flux with both nearest neighbor (NN) and isotropic next NN (NNN) hoppings through numerical calculation\cite{APT_Thouless_1994}, which exhibited three phases with the extended, localized and critical states respectively. Later, Avila and collaborators obtained the phase diagram of this model with NN and general  NNN  hoppings rigorously in his Fields medal work by means of the transfer matrix, which is now called Avila's global theory \cite{Avila2015,Jitomirskaya2012,SquareNNN_2017}. Recently the 1D incommensurate Aubry–André–Harper (AAH) model
reduced from the square-lattice Hofstadter model with irrational flux has attracted a lot of interest because of its application in quasicrystals \cite{Quasicrystal}, and a renormalization group (RG) theory was successfully applied to this system and its extended models in studying the phase diagram and phase transitions in such systems\cite{RG,RG2}.

\begin{figure*}[htbp]  % figure* = 跨双栏；htbp 适配双栏排版
    \centering
    % \textwidth = 整页宽度（占满双栏）
    \includegraphics[scale=1.6]{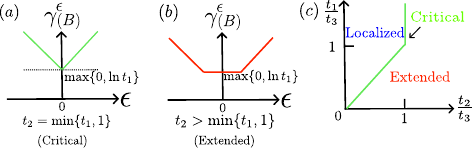}  
    \caption{\justifying (a) and (b) show schematically  the behavior of $\gamma^\epsilon(B)$ as a function of $\epsilon$  in the regime $t_2\ge \min\{t_1,1\}$, for which $\gamma^{\epsilon\to 0}(T)=0$ ($t_3$ is set to $1$). (a)$\gamma^\epsilon(B)$ as a function of $\epsilon$ in the case $t_2=\min\{t_1,1\}$.  The derivative of $\gamma^\epsilon(B)$ at $\epsilon\to 0$ is discontinuous, indicating critical states in this case. (b)$\gamma^\epsilon(B)$ as a function of $\epsilon$ in the case $t_2>\min\{t_1,1\}$. The derivative of $\gamma^\epsilon(B)$ at $\epsilon\to 0$ is zero, indicating extended states in this regime. (c) The localization phase diagram of the hexagonal-lattice Hofstadter model with a \textcolor{blue}{Diophantine} irrational magnetic flux from Avila's global theory ($t_3$ is restored in the diagram and $t_1/t_3 \neq 1$).  The green line corresponds to critical states. The left region of the green line corresponds to localized states and the right region of the green line corresponds to extended states.}\label{GammaB}
\end{figure*}

Though the theoretical studies on the irrational flux Hofsatdter model were carried out mostly on the square lattice, a lot of Hofstadter models achieved in experiments in recent years, such as in Moire lattice \cite{Moire_Lattice}, graphene superlattice  \cite{graphene_superlattice}, heterorstructures \cite{Heterorstructures}, have hexagonal lattice structure, for which the irrational flux case has been barely investigated. One of the main difference between the hexagonal and square lattice is that the former has two sublattices. In the most general case, this makes the transfer matrix of  Avila's global theory a $4*4$ matrix in the hexagonal lattice instead of $2*2$ matrix in the square lattice. The general $4*4$ transfer matrix in the irrational flux case does not guarantee exact solubility. However, we show in this work that with only NN hopping in the hexagonal lattice, the system can be characterized by a  $2*2$ transfer matrix and is exactly solvable. We then obtained the phase diagram and critical properties of the irrational flux hexagonal lattice with NN hopping exactly by Avila's global theory, as shown in Fig.\ref{GammaB}(c).

We  applied the RG theory to verify the results from Avila's global theory of the hexagonal lattice. The RG theory can determine the localized regime and part of the extended regime exactly, but not the critical regime and the other part of extended regime. For these regimes, we then computed the fractal dimension numerically to determine the phases. Both the results from the RG and numerical analysis are consistent with what we obtained from Avila's global theory. 

We compared our results of the hexagonal lattice with irrational flux with the recently studied exactly solvable spin-$1/2$ quasiperiodic (QP) models in Ref.\cite{Unification}. Our model is exactly solvable but it does not satisfy all the conditions of exact solubility proposed in Ref.\cite{Unification}. Moreover, the critical states in our model do not correspond to the existence of generalized incommensurate zeroes in the hopping matrix, in contrast to the proposal in Ref.\cite{Unification}. Our model in this work is then a different type of exactly solvable QP model from that in Ref.\cite{Unification}.

\begin{figure}[t]
    \centering  \includegraphics[scale=1.3]{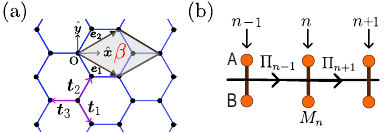}
     \caption{
   \justifying
(a)Hexagonal lattice with nearest-neighbor hoppings $\{t_1,t_2,t_3\}$ where $\boldsymbol{e}_1, \boldsymbol{e}_2$ are the  primitive lattice vectors. The gray shaded unit cell encloses  a magnetic  flux $\phi=\boldsymbol{B}\cdot(\boldsymbol{e}_1\times\boldsymbol{e}_2)= 2\pi\beta$ with $\beta$ irrational. (b) 1D quasiperiodic chain of the hexagonal lattice Hofstadter model after Fourer transformation in $\boldsymbol{e}_1$ direction. Each site has two sublattices $A$ and $B$ in the chain.}
\label{fig:Lattice}   
\end{figure}

{\it Model.}  
We study a hexagonal lattice   in a perpendicular magnetic field $\boldsymbol{B}=\partial_x A_y-\partial_y A_x$ with the vector potential $\mathbf{A}=\hat{y}(x+\sqrt{3} y)B$, as shown in Fig.\ref{fig:Lattice}(a). The unit vectors of the primitive lattice at zero magnetic field are $\boldsymbol{e}_1=(\frac{3}{2}\hat{x}-\frac{\sqrt{3}}{2}\hat{y})a$ and $\boldsymbol{e}_2=(\frac{3}{2}\hat{x}+ \frac{\sqrt{3}}{2}\hat{y})a$ with $a$ the bond length and $\hat{x}, \hat{y}$ the unit vectors in the $x$ and $y$ directions. The magnetic flux in each unit cell  is $\phi=\boldsymbol{B}\cdot(\boldsymbol{e}_1\times\boldsymbol{e}_2)= 2\pi\beta,$ where $\beta$ is irrational. Hereafter we set $a=\hbar=e=1$.

With the above choice of the vector potential $\mathbf{A}$, the tight binding Hamiltonian of the hexagonal lattice with irrational flux is periodic in $\mathbf{e}_1$ direction but not in $\mathbf{e}_2$ direction. We  assume periodic boundary condition in the $\boldsymbol{e}_1$ direction and open boundary condition in $\boldsymbol{e}_2$ direction, and study the localization behavior of the electron wavefunction in $\boldsymbol{e}_2$ direction. After a Fourier transformation in $\boldsymbol{e}_1$ direction,
 the tight-binding Hamiltonian of a hexagonal lattice in the magnetic field with NN hopping takes the following form \cite{VHS,Gao_Chen_2026} 
\begin{align}   
H&=-\sum\limits_{n}\Big(t_3a_{n}^{\dagger}b_{n}+t_1e^{-i(k_1+2\pi n\beta)}a_{n}^{\dagger}b_{n}\notag\\ &+t_2e^{i2\pi n\beta }a_{n}^{\dagger}b_{n+1}\Big)+h.c. \label{Hexagonal}
\end{align} 
where $a^\dag_{n}, b^\dag_{n}$ are spin-polarized electron creation operators on site $(n,A)$ and $(n,B)$ with $n\in \mathbb{Z}$, and $t_i, i=1, 2, 3$ are the NN hopping parameters in Fig.\ref{fig:Lattice}(a). 
We define the state on site $n$ as $\psi_{n}\equiv \begin{pmatrix}
a_{n}^{\dagger}\ket{0}\\    b_{n}^{\dagger}\ket{0}
\end{pmatrix},$ and denote its components by 
    $\psi_n^{A}\equiv a_{n}^{\dagger}\ket{0},\ 
    \psi_{n}^B\equiv b_{n}^{\dagger}\ket{0}$.

The Hamiltonian Eq.(\ref{Hexagonal}) can be expressed as
\begin{equation}\label{Hamiltonian_2}
 H\psi_{n}=M_{n}\psi_{n}+\Pi_{n} \psi_{n+1}+\Pi_{n-1}^{\dagger}\psi_{n-1}, 
\end{equation} 
where the on-site potential matrix $M_n$ and hopping matrix $\Pi_{n}$ are given by
\begin{align}
    M_{n}&=-\begin{pmatrix}
        0 & t_3+t_1e^{i(k_1+2\pi n\beta)}\\
        t_3+t_1e^{-i(k_1+2\pi n\beta)}&0
\end{pmatrix},\\ 
     \Pi_{n}&=-\begin{pmatrix}
        0 & t_2e^{-i2\pi n \beta}\\
        0&0    \end{pmatrix}.
\end{align} 
The Hamiltonian Eq.(\ref{Hamiltonian_2}) has chiral symmetry which is characterized by $\{\mathcal{C},H\}=0$
with the chiral symmetry operator $\mathcal{C}=\mathbf{I}_{N}\otimes \sigma_{z}$, where $N$ is the number of unit cells in $\boldsymbol{e}_2$ direction.

{\it Avila's global theory.} The  eigenvalue equations of Hamiltonian Eq.(\ref{Hamiltonian_2}) read  
\begin{align}
    -E\psi_{n}^{A}&=\bar{c}_n(k_1)\psi_{n}^{B}+t_2e^{-i2\pi n\beta}\psi_{n+1}^{B},\label{psiB} \\ 
    -E\psi_n^{B}&=c_{n}(k_1)\psi^{A}_{n}+t_2e^{i2\pi(n-1)\beta}\psi_{n-1}^{A},\label{psiA}
\end{align} 
where $ c_{n}(k_1)=t_3+t_1e^{-i(k_1+2\pi n\beta)}$ and $\bar{c}_{n}(k_1)$ denotes the complex conjugate of $c_n(k_1).$ From Eq.(\ref{psiA}), we can express $\psi^A_n$ by $\psi_n^B$ and $\psi^A_{n-1}$. After substituting $\psi^A_n$ to Eq.(\ref{psiB}), we can get an iterative relation for the vector $(\psi_{n}^{B},\psi_{n-1}^{A})^{\mathrm{T}}$ as
\begin{equation}
c_{n}(k_1)\begin{pmatrix}
        \psi_{n+1}^{B}\\
        \psi_{n}^{A}
\end{pmatrix}=B_{n}(k_1)\begin{pmatrix}
        \psi_{n}^{B}\\
        \psi_{n-1}^{A}
    \end{pmatrix},
\end{equation}
where
\begin{equation}
B_{n}(k_1)=\begin{pmatrix}
\frac{(-\vert c_{n}(k_1)\vert^2+E^2)e^{i2\pi n\beta}}{t_2}  & Ee^{2i\pi(2n-1)\beta}\\
-E &-t_2e^{i2\pi(n-1)\beta}  \end{pmatrix}.
\end{equation} For simplicity, we set $t_3=1$ as the unit of the hopping parameters and $t_1\neq 1$ such that $c_{n}(k_1)\neq 0.$ 
We then get a $2\times 2$ transfer matrix of the system as 
\begin{align}
    T_{n}(k_1)= \frac{1}{c_{n}(k_1)}B_{n}(k_1),\ B_n(k_1)\in GL(2,\mathbb{C}).
\end{align}

With the transfer matrix $T_{n}(k_1)$, we can use Avila's global theory to study the localization behavior of the states. Avila's global theory enables the analytic calculation of the Lyapunov exponent (LE) from the transfer matrix \cite{Avila2015,Jitomirskaya2012,SquareNNN_2017}, which is the inverse of the localization length $\xi$ under a certain condition, as shown in Appendix A.  
The LE can be obtained from the transfer matrix $T_{n}(k_1)$ as \cite{Avila2015,Jitomirskaya2012,SquareNNN_2017}
\begin{align}
\gamma(T)&=\lim\limits_{n\to \infty}\frac{1}{2\pi n}\int_{0}^{2\pi} \ln \vert\vert \prod\limits_{j=1}^{n}\frac{B_{j}(k_1)}{c_{j}(k_1)}\vert\vert d k_1,\notag\\  
&=\gamma(B)-I,\label{gamma}
\end{align}
where 
\begin{align}
\gamma(B)\equiv \lim\limits_{n\to \infty}\frac{1}{2\pi n}\int_{0}^{2\pi}\ln \vert\vert\prod\limits_{j=1}^{n}B_{j}(k_1)\vert\vert dk_1
\end{align}
is the LE of the matrix $B_{n}(k_1)$ and 
\begin{equation}\label{I_integral}
I\equiv \frac{1}{2\pi}\int_{0}^{2\pi} \ln \vert c_{j}(k_1)\vert dk_1. 
\end{equation}
 As shown in Appendix B, the 
LE $\gamma(T)\geq 0$. When $\gamma(T)> 0$, the localization length is finite and the corresponding state is localized. When $\gamma(T)=0$, the state is either extended or critical.

The second term $I$ of Eq.(\ref{gamma}) can be calculated directly through the Jensen's formula. The coefficient $\gamma(B)$ can be calculated by first complexifying $\gamma(B)$ through 
\cite{Avila2015,Jitomirskaya2012,SquareNNN_2017},
\begin{equation}
k_1\to k_1+2\pi i\epsilon,\ \gamma(B)\to \gamma^{\epsilon}(B)\notag,\ \epsilon\in \mathbb{R}
\end{equation} 
and then taking the limit $\gamma(B)=\lim_{\epsilon\to 0} \gamma^{\epsilon}(B)$ at the end. 

The asymptotic behavior of  $\gamma^{\epsilon}(B)$ at $\epsilon\to \pm \infty$ can be easily obtained through 
 \begin{equation}
 \prod\limits_{j=1}^{n}B_{j}(k_1+2\pi i\epsilon) \overset{\epsilon\to \pm \infty}{=\joinrel=\joinrel=}\begin{pmatrix}
        (-\frac{t_1}{t_2})^n e^{\pm 2\pi n\epsilon}  e^{\mp i n k_1}&0\\
        0&0
     \end{pmatrix}+o(1), 
 \end{equation}
where $o(1)/e^{|\epsilon|}$ is infinitesimal. We then obtain 
 \begin{equation}
 \gamma^{\epsilon\to \pm \infty}(B)  =\ln(\frac{t_1}{t_2})\pm 2\pi\epsilon.
 \end{equation}

The limit of $\gamma^{\epsilon\to 0}(B)$ can be obtained from $\gamma^{\epsilon\to \pm \infty}(B)$ through extrapolation due to the following properties of $\gamma^{\epsilon}(B)$ for any Diophantine irrational number $\beta$ \cite{Avila2015,Jitomirskaya2012,SquareNNN_2017}: (1)$\gamma^{\epsilon}(B)$ is convex; (2)continuous; (3)the right derivative $\lim\limits_{h\to 0+} \frac{\gamma^{\epsilon+h}(B)-\gamma^\epsilon(B)}{h}=2\pi \mathbb{Z}$ for all range of $\epsilon$;\ (4)$\gamma^{\epsilon\to 0}(B)\geq I^{\epsilon\to 0}$ (as shown in the Appendix); and (5)if $\gamma^{\epsilon\to 0}(T)>0$,  then $\lim\limits_{\epsilon\to0}d \gamma^\epsilon(B)/d\epsilon$ is not well-defined. In this work, we only consider the Diophantine irrational flux case. Since most known irrational numbers are Diophantine \cite{Note}, the above properties are valid for the system with most irrational flux $\beta$. So are the results we obtained in this work. 

We note that after the complexification, $c_j(k_1+2\pi i\epsilon)$ has zero points at $t_1 e^{2\pi\epsilon}=1$. For the reason, $\gamma^\epsilon(T)$ is not analytic at this $\epsilon$ and the convexity at this $\epsilon$ is lost for $\gamma^\epsilon(T)$. We then can not get $\gamma^\epsilon(T)$ at $\epsilon\to 0$ by extrapolation from $\gamma^{\epsilon\to\pm \infty}(T)$. However, since $\gamma^{\epsilon}(B)$ is still well-behaved, we can get $\gamma^{\epsilon\to 0}(B)$ by extrapolation from $\gamma^{\epsilon\to\pm \infty}(B)$ and then add $-I^\epsilon$ to get $\gamma^\epsilon(T)$.

We then get 
\begin{equation}\label{gammaB} 
\gamma^{\epsilon}(B)=\max\{\ln(\frac{t_1}{t_2})+2\pi|\epsilon|, I^{\epsilon\to 0} \}.
\end{equation}

The second term of Eq.(\ref{gamma}) after complexification is obtained from the Jensen's formula as (see Appendix C)
\begin{align}
 I^\epsilon & = \frac{1}{2\pi}\int  \ln\vert c_{j}(k_1+2\pi i\epsilon)\vert d k_1   \nonumber\\
 &=\begin{cases}
       \ln t_1 +2\pi \epsilon,& \text{if}\ \  t_1e^{2\pi \epsilon}\geq 1,\\
        0, &\text{if}\ \ t_1e^{2\pi\epsilon}<1.
    \end{cases}\label{Jensen}
\end{align}
    
Since $\gamma^{\epsilon\to 0}(T)$ can not be negative (see Appendix B), we get $\gamma^\epsilon(T)$ at $\epsilon\to 0$ 
by combining Eqs. (\ref{gammaB}) and (\ref{Jensen}) as
\begin{align} 
\gamma(T) &=\begin{cases}
        \ln(\frac{1}{t_2}), & \text{if} \  t_1>1>t_2,\\ 
        \ln(\frac{t_1}{t_2}), & \text{if}\ 1>t_1>t_2,\\
        0, & \text{if}\ t_2\ge \min\{t_1,1\}.
    \end{cases}\label{gamma_T}
\end{align}

In our system, the LE $\gamma(T)$ corresponds to the inverse of localization length (see Appendix A). 
From Eq.(\ref{gamma_T}), we can see that at $t_2<{\rm min}\{t_1, 1\}$, $\gamma(T)>0$ and the states are localized with localization length $\xi=1/\ln(\frac{\min(t_1, 1)}{t_2})$.

When the LE satisfies $\gamma(T)=0$, the correlation length goes to infinity. According to Ref.\cite{Jitomirskaya2012}, 
if the derivative of $\gamma^{\epsilon}(B)$ at $\epsilon\to 0$ is continuous and equal to zero, 
the state is extended. If the derivative of $\gamma^{\epsilon}(B)$ is discontinuous at $\epsilon\to 0$,
the state is critical.

 From Eq.(\ref{gamma_T}), at $t_2\geq{\rm min}\{t_1, 1\}$, $\gamma(T)=0$. 
To distinguish the extended and critical states in this regime, we need to compute $\lim\limits_{\epsilon\to0}d \gamma^\epsilon(B)/d\epsilon$.   
From Eq.(\ref{gammaB})-(\ref{gamma_T}), we get: 

(1)At $t_1=t_2<1$, the LE index $\gamma^{\epsilon}(B)$ in the neighborhood of $\epsilon = 0$ behaves as 
\begin{equation}
    \gamma^{\epsilon}(B)=
        2\pi \vert\epsilon\vert, \ \text{if}\  \epsilon\in\mathbb{R}   \label{critical_1}.
\end{equation}
The left and right derivatives of $\gamma^{\epsilon}(B)$ at $\epsilon\to 0$ are not equal so the states are critical in this regime.

(2)At $t_1>t_2=1$, 
the LE index $\gamma^{\epsilon}(B)$ in the neighborhood of $\epsilon = 0$ behaves as
\begin{equation}
    \gamma^{\epsilon}(B)=\ln(t_1)+2\pi\vert\epsilon\vert,\ \mathrm{if}\ \epsilon\in\mathbb{R}.
\end{equation}
For the same reason as in case $(1)$, the states in this regime are critical too. 

(3)At $t_2>t_1$, the LE index $\gamma^{\epsilon}(B)$ in the neighborhood of $\epsilon = 0$ behaves as
\begin{equation}
\gamma^{\epsilon}(B)=\max\{0,\ln t_1\},\ {\rm at }\ \epsilon\in(-\frac{1}{2\pi}\ln \frac{t_2}{t_1},\frac{1}{2\pi}\ln \frac{t_2}{t_1})\\.
\end{equation}
The derivative of $\gamma^\epsilon (B)$ at $\epsilon\to 0$ is continuous and  zero.  For the reason, the states are extended in this regime. 

(4)At $t_1> t_2>1$, the LE index $\gamma^{\epsilon}(B)$ in the neighborhood of $\epsilon = 0$ behaves as
\begin{equation}
\gamma^{\epsilon}(B)=\ln t_1,\ {\rm at }\ \epsilon\in(-\frac{1}{2\pi}\ln t_2,\frac{1}{2\pi}\ln t_2)\\.
\end{equation}
The derivative of $\gamma^\epsilon (B)$ at $\epsilon\to 0$ is also continuous and zero so the states in this regime are extended too.

From the above analysis, we get the behavior of $\gamma^{\epsilon}(B)$ at $\gamma^{\epsilon\to 0}(T)=0$ in the regime $t_2\ge \min\{t_1,1\}$ as summarized in Fig.\ref{GammaB}(a)-(b). We can see that at $t_2=\min\{t_1,1\}$, the states are critical as indicated by Fig.\ref{GammaB}(a), and at $t_2 > \min\{t_1,1\}$, the states are extended
as indicated by Fig.\ref{GammaB}(b).

Combing the localized, critical and extended regimes from the above analysis, we get the exact localization phase diagram of the hexagonal Hofstadter model  with Diophantine irrational flux from Avila's global theory as shown in Fig.\ref{GammaB}(c) (where $t_3$ is restored). We can see that all the regimes have pure phases and there is no mobility edge. The phase diagram is independent of the value of $k_1$ and the Diophantine irrational flux number $\beta$.

Near the phase transition from the localized state to the critical state, the correlation length $\xi=1/\ln(\frac{\rm min(t_1, 1)}{t_2})$ which diverges
as $\xi\sim \delta^{-1}$ where $\delta\equiv \rm min(t_1, 1)-t_2 \to 0 $. The critical index $\nu=1$,
which is the same as for the phase transition from the  localized to extended states of the AAH model\cite{RG}.

{\it RG theory.} We next compare the above results with those from the RG theory.

In the RG theory, the irrational flux is approached by a series of  commensurate approximations (CA) \cite{RG,RG2}. For example, one can use the Fibonacci numbers \cite{APT_Thouless_1994} to approach the irrational flux $\beta=\frac{\sqrt{5}-1}{2}$ as $\beta=\lim\limits_{n\to \infty}p_{n}/q_n=\lim\limits_{n\to \infty} F_{n-1}/F_{n}$, where $F_{n}$ is the $n$-th Fibonacci number and $F_{n}$ and $F_{n-1}$ are coprime. The length of the magnetic unit cell (UC) in $\boldsymbol{e}_2$ direction corresponding to flux $p_{n}/q_n$ is $L=q_n$, which increases with the increase of $n$. The behavior of the commensurate case at $n$ or $L \to \infty$ gives the properties of the irrational flux case.

To get the renormalized couplings of the $n$-th order CA, we compute the characteristic polynomial $\mathcal{P}_n(E)\equiv \det(H-E\mathbf{I})$ which determines the energy spectrum when setting $\mathcal{P}_n(E)=0$. 
The polynomial $\mathcal{P}_n(E)$ for the hexagonal lattice with NN hopping and flux $\phi_n=p_n/q_n=p_n/L$ per (primitive) UC is \cite{RG,RG2}
\begin{equation}
    \mathcal{P}_n(E)=G_n(E)-\xi_n (k_1,k_2),
\end{equation} 
where $G_n(E)$ is an even function of energy due to the chiral symmetry of the Hamiltonian and 
\begin{eqnarray}
\xi_n( k_1,k_2)&=&\vert t_1^{L} e^{i k_1 L-i\pi(L-1)}+t_2^{L}e^{ik_2}+t_3^L\vert \nonumber\\
& =&\sqrt{t_1^{2L}+t_2^{2L}+t_3^{2L}+\eta_L(k_1,k_2)}
\end{eqnarray}
with 
\begin{align} 
\eta_L (k_1,k_2)&=2(t_2t_3)^{L}\cos(k_2)
-2(-1)^{L}(t_1 t_3)^{L}\cos(k_1 L)\notag\\ 
 & -2(-1)^{L}(t_1t_2)^{L}\cos(k_2- k_1 L).
\end{align}
For convenience, we set $t_3=1$ and $\eta_L (k_1,k_2)$ becomes
\begin{align}
    \eta_L(k_1,k_2)&= 2t_2^L\cos(k_2)-2(-1)^{L}t_1^L\cos(k_1 L)\notag\\ 
&-2(-1)^{L}(t_1t_2)^{L}\cos(k_2-k_1 L).
\end{align}

We can get the momentum dependence of the states from the analysis of $\eta_L(k_1,k_2)$.
At $n$ or $L \to \infty$, if $t_1>t_2$ and $t_2<1$, 
%then $2t_2^{L}\cos(k_2)\to 0 $ and $ \vert 2(-1)^{L}t_1^L\cos(k_1 L)\vert \gg \vert 2(t_1t_2)^L\cos(k_2-k_1 L)\vert$. 
the most relevant term in $\eta_L(k_1,k_2)$ is the hopping term  $-2(-1)^{L}t_{1}^L\cos(k_1 L)$ in $\boldsymbol{e}_1$ direction with the $n$-th order renormalized hopping $\sim 2 t^L_1$. The hopping in $\boldsymbol{e}_2$ direction is irrelevant and 
for a given energy, $k_2$ can take any value.  This means that the electrons are  delocalized in momentum space along the  $\boldsymbol{e}_2$ direction and localized in the real space along $\boldsymbol{e}_2$ direction. 
On the other hand, if $t_1<t_2$ and $t_1<1$, at $L\to \infty$, the most relevant term in $\eta_L(k_1,k_2)$ is the hopping term in $\boldsymbol{e}_2$ direction, i.e.,  $2t_2^L\cos(k_2).$ This means that $k_2$ takes a specific value for a given energy $E$, corresponding to a localized state in momentum space but a delocalized state in real space along  $\boldsymbol{e}_2$ direction. Since the energy part $G(E)$ is separated from the momentum part $\xi(k_1, k_2)$ in $\mathcal{P}(E)$, there is no mobility edge in the system. 

The analysis of the above two regimes from the RG theory is consistent with the phase diagram obtained from Avila's global theory in Fig.\ref{GammaB}(c). However, the RG theory can not determine the phase of the above model beyond these two regimes. For the remaining regimes, we can compute the fractal dimension (FD) numerically to confirm the results from Avila's global theory.

\begin{figure}
\includegraphics[scale=0.8]{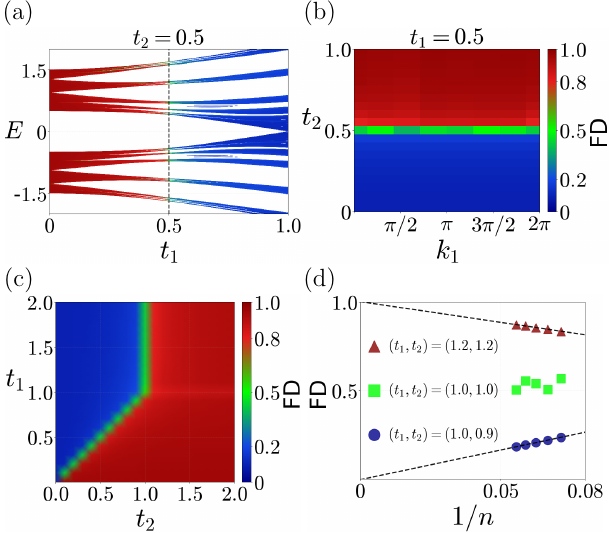}
\caption{\justifying Numerical results from the FD analysis of the hexagonal lattice with irrational flux $\beta=\frac{\sqrt{5}-1}{2}$, $t_3=1$ and $N=F_{16}=987$ \cite{Data}. (a)The FD of the eigenstates as a function of $E$ and $t_1$ at $t_2=0.5$. The FD of the eigenstates is independent of energy, indicating that the system has no mobility edges. (b)The FD of the ground states as a function of  $k_1$ and $t_2$  at $t_1=0.5$. The phase diagram does not depend on the momentum $k_1$ along $\boldsymbol{e}_1$ direction. (c)Localization phase diagram of the ground states for $\{t_1,t_2\}\in(0, 2)$. The blue, green and red regimes correspond to the localized, critical and extended phases respectively. (d)Extrapolation of FD to $n\to \infty$ for a localized (blue), critical (green) and extended (red) state in (c) respectively, where $n$ is the subscript of the Fibonacci number $F_n$. }\label{FD_analysis}
\end{figure}

{\it FD Analysis}.
 Numerically the extended, localized and critical states can be distinguished by computing the FD of the states \cite{Anomalous_mobility_edges,MosaicModel_2020,Critical_Phases,Unification}.  
Assuming that the wavefunction of the Hamiltonian Eq.(\ref{Hamiltonian_2}) at site $j$ in the $\boldsymbol{e}_2$ direction  is $u_{j}^{E}$ for the state with given energy $E$ and $k_1$, the FD of the state can be defined as \cite{Anomalous_mobility_edges,MosaicModel_2020,Critical_Phases,Unification}
   \begin{equation}
       \mathrm{FD}\equiv -\lim\limits_{N\to +\infty}\frac{\ln(\sum\limits_{j=1}^{N}\vert u_{j}^{E}\vert^{4})}{\ln (N)},\ {\rm with}\ \sum\limits_{j=1}^{+\infty}\vert u_{j}^{E}\vert^2=1.
   \end{equation}

 For the localized state, the wavefunction is mainly distributed  in  a finite region, therefore FD=0. For the extended state, the wavefunction is distributed uniformly in each UC, i.e., $ \vert u_{j}^{E}\vert \approx \frac{1}{\sqrt{N}}$. For the reason, $\ln(\sum\limits_{j=1}^{N}\vert u_{j}^{E}\vert^{4})=-\ln(N)$ and FD=1. For the critical state, FD is in between, i.e., $0<\mathrm{FD}<1$.

We calculated the FD of the eigenstates of the hexagonal lattice with irrational flux $\beta=\frac{\sqrt{5}-1}{2}$, still employing the Fibonacci numbers $F_{n-1}/F_n$ to approach the irrational flux in the calculation. We set the number of sites in $\boldsymbol{e}_2$ direction to be $N=F_{n}$ and solve the Hamiltonian matrix of Eq.(\ref{Hamiltonian_2}) for a given $k_1$ in real space numerically. At very large $N$, the result approaches that of the irrational flux $\beta=\frac{\sqrt{5}-1}{2}$. From the solved wavefunction, we can compute the FD of the state.

The results of the FD  analysis  are shown in Fig.\ref{FD_analysis} \cite{Data}. Figure \ref{FD_analysis}(a) shows that the FD of the eigenstates is independent of energy, i.e.,
 there is no mobility edge for the system. We can then choose to compute the FD of the ground states.
 Figure \ref{FD_analysis}(b) shows that FD is independent of momentum $k_1$, consistent with the result from  Avila's global theory. We then fixed $k_1$ and set $t_3=1$, and computed the FD of the ground states of the system for $t_1, t_2\in (0, 2)$. The results are shown in the phase diagram in Fig. \ref{FD_analysis}(c). The blue and red regimes are localized and extended states respectively, and the green line corresponds to critical states.  Figure \ref{FD_analysis}(d) shows the extrapolation of the FD to $n\to \infty$, i.e., the irrational flux limit, for states in the red, green and blue regimes. The extrapolation shows that the FD in the red (blue) regime goes to $1 (0)$ at $n\to \infty$, corresponding to an extended (localized) state in the irrational flux limit. The FD of the state on the green line remains about $0.5$ during the extrapolation, characteristic of a critical state.
  The phase diagram from the numerical analysis of FD shown in Fig. \ref{FD_analysis}(c) is then 
  completely consistent with the exact phase diagram we obtained from Avila's global theory.

{\it Discussions.} Recently, a type of one-dimensional spin-$1/2$ QP model with the Hamiltonian of the form of Eq.(\ref{Hamiltonian_2}) was investigated in Ref.\cite{Unification}, for which the elements of hopping matrix $\Pi_n$ and potential matrix $M_n$ are either uniform or QP. The condition of the existence of mobility edge, critical states and exact solubility was discussed for the model. 
Some of these discussions apply to our model in this work. For example, our model has chiral symmetry and as a consequence there is no mobility edge.
Also in our model ${\rm Det}\ \Pi_n=0$  which is consistent with one of the conditions of exact solubility proposed in Ref.\cite{Unification}. But the potential matrix in our model does not satisfy their second condition that the off-diagonal matrix elements are either constant or pure QP. 
Moreover, the critical states in our model do not correspond to the existence of generalized incommensurate zeroes in the hopping matrix, as opposed to the proposal in Ref.\cite{Unification}. Our model in this work then provides a different type of exactly solvable QP model from that in Ref.\cite{Unification}.

This work  focuses on the non-interacting hexagonal Hofstadter model with irrational flux. It has been known that interaction can significantly change the ground state of the non-interacting quasi-periodic AAH model originated from the square lattice Hofstadter model with irrational flux \cite{MBL_Numberical,MBL_Experiment,MBL_Critical}. For example, a single-particle localized state in the AAH model may evolve to a many-body localized state or a many-body ergodic delocalized state depending on the strength of the interaction, whereas the single-particle extended state usually evolves to a many-body ergodic delocalized state under interaction. Whether these conclusions may be generalized to the hexagonal lattice remains to be explored in future work.

The Hofstadter model in hexagonal lattice can be experimentally realized in superlattices with nanometer scale UCs, such as Moire bilayer graphene \cite{Moire_Lattice}, or graphene on a properly aligned Boron-Nitride substrate \cite{Moire_Lattice,graphene_superlattice,Heterorstructures}. It can also be simulated by cold atoms in optical lattices shown in Refs.\cite{Experiment1,Experiment2}, or photonic crystals in Refs.\cite{Experiment3,Experiment4}. These systems provide a variety of promising platforms to verify the theoretical results of the hexagonal Hofstadter model in this work.

{\it Summary.} In summary, we studied the localization phase diagram of the hexagonal lattice with Diophantine irrational flux. With only NN hopping, the system can be characterized by a $2*2$ transfer matrix and the phase diagram can be obtained exactly by Avila's global theory. We also applied the RG theory to study the model, which can determine part of the phase diagram. Finally, we studied the model by the fractal dimension analysis numerically. Both the results from the RG and fractal dimension analysis are consistent with what we obtained from Avila's global theory. Our results can be tested in various recently achieved hexagonal Hofstadter models in experiments. 
 
{\it Acknowledgment.} We thank Yiqian Wang, Jiahao Xu, Xiangshuo Kong  for very helpful discussion. This work is supported by the National NSF of China under Grant No. 11974166 and No. 12271245, the NSF of Jiangsu Province under Grant No.BK20231398 and the Nanjing University International Research Seed Fund.

\appendix

\begin{widetext}

\section{Relationship between the Lyapunov exponent and the localization length}

In this Appendix, we present the relationship between the LE and the localization length.

We first note that  the definition of the LE \( \gamma(T) \) given in Eq.(\ref{gamma}) in the main text is equivalent to the following definition we use in the Appendix\cite{Jitomirskaya2012}:
\begin{equation}\label{gamma_definition}
\gamma(T) \equiv \lim\limits_{n\to \infty}\frac{1}{2\pi n}\int_{0}^{2\pi} \ln \vert\vert \prod\limits_{j=1}^{n}T_{j}(k_1)\vert\vert d k_1\equiv\lim\limits_{n\to \infty}\frac{1}{n}\ln \vert\vert\prod\limits_{j=1}^{n}T_{j}(k_1) \vert\vert.
\end{equation}
Note that $\gamma(T)$ is independent of $k_1$ in the second definition in Eq.(\ref{gamma_definition}) after the sum of the logs over all the steps from $j=1$ to $j\to \infty$ for irrational flux $\beta$. For the reason, the average of $\gamma(T)$ over a period of $k_1$ in the first definition is  equivalent to the second definition.

To get the relationship between the LE and the localization length, we introduce the Oseledets Theorem:

\textbf{Oseledets Theorem} (see Appendix G of Ref.\cite{Jitomirskaya2012}): Let $(A_{n})_{n\in \mathbb{N}}$ be a sequence of $2*2$ matrices in $GL(2,\mathbb{C}).$ Set $d_{n}:=(\det A_{n})^{1/2},\ D_{n}:=\frac{1}{d_{n}}A_{n},\ n\in\mathbb{N}$, then $ D_{n}\in SL(2,\mathbb{C})$, i.e., $\det D_n=1$. Suppose that $m:=\lim\limits_{n\to \infty}\frac{1}{n}\sum\limits_{k=1}^{n}\ln \vert d_{k}\vert$ exists and is finite, and $\lim\limits_{n\to \infty}\frac{1}{n}\ln \vert\vert A_{n}\vert\vert=0$ (i.e., $A_{n}$ is bounded), and  $\gamma(A):=\lim\limits_{n\to \infty}\frac{1}{n}\ln \vert\vert A_{n}\cdots A_{1}\vert\vert$ exists and is finite with $\gamma(D)=\gamma(A)-m> 0$, then there is a one-dimensional subspace $S\subset \mathbb{C}^{2}$ such that for a vector $v\in \mathbb{C}^{2}$,  
\begin{itemize}
    \item (i): $\frac{1}{n}\ln\vert\vert A_{n}\cdots A_{1}v\vert\vert  \xrightarrow{n\to\infty} -\gamma(A)+2m$ if $v\in S$,
 \item (ii): $\frac{1}{n}\ln \vert\vert A_{n}\cdots A_{1}v\vert\vert \xrightarrow{n\to\infty} \gamma(A)$ if $v\in \mathbb{C}^2/S$.
\end{itemize} 

Case (i) corresponds to an exponentially decaying state generated by the transfer matrix $D\equiv\lim\limits_{n\to \infty}D_n\cdots D_1$ and case (ii) an exponentially increasing state. For the reason, case (i) is the physical state we keep.

Suppose $A_n$ is the transfer matrix of a physical system, i.e., 
 $\psi_{n}\equiv A_{n}\cdots A_{1}\psi_{0}$, where  $\psi_{0}$ and $\psi_n$ are the states on site $0$ and $n$ respectively, 
 for $A_n\in GL(2,\mathbb{C})$, there exists a state $\psi_n$ such that 
\begin{equation}
    \frac{1}{n}\ln |\vert \psi_{n} \vert|\xrightarrow{n\to\infty} -\gamma(A)+2m \Rightarrow |\vert\psi_{n}\vert| \sim e^{-(\gamma(A)-2m)n}.
\end{equation}
The localization length of the state $\psi_n$ is $\xi=1/(\gamma(A)-2m)$, where $\gamma(A)$ is the LE of matrix $A$. In the case $m=0$, $\xi=1/\gamma(A)$.

\section{Proof of the non-negativity of Lyapunov exponent}
In this appendix, we prove that the LE index $\gamma^{\epsilon\to 0}(T)$ in our model is non-negative.

We first prove that 
for any transfer matrix \(D_n\in SL(2,\mathbb{C})\), its LE  \(\gamma(D)=\lim\limits_{n\to \infty }\frac{1}{n}\ln\vert\vert D_{n}\cdots D_{1}\vert\vert\) satisfies \(\gamma(D)\ge 0\).
 \begin{proof}
Suppose the eigenvalues of $D^{\dagger}D$ are $\lambda_1(D), \lambda_2(D)$ and $\lambda_1(D)\ge \lambda_2(D)\ge 0$, the singular values of the matrix $D$ are then 
\begin{equation}
(\sigma_1(D), \sigma_2(D))=(\sqrt{\lambda_1}, \sqrt{\lambda_2}),\ \sigma_1(D)\ge \sigma_2(D)\ge 0,\ \det(D)=\sigma_{1}(D)\sigma_{2}(D)=1,
\end{equation}
which indicates that $\sigma_1(D)\geq 1 \geq \sigma_2(D)\geq 0$.
We then adopt the spectral norm of the matrix, which yields
\begin{equation}
   \gamma(D)=\lim\limits_{n\to \infty}\frac{1}{n}\ln\vert\vert D_{n}\cdots D_{1}\vert\vert=\lim\limits_{n\to\infty}\frac{1}{n}\ln \sigma_{1}(D)\ge 0. 
\end{equation}   
\end{proof}

We then consider the transfer matrix in the hexagonal Hofstadter model in this work, which is $GL(2,\mathbb{C})$ and takes the form 
\begin{equation}
     T_{n}(k_1)= \frac{1}{c_{n}(k_1)}\begin{pmatrix}
\frac{(-\vert c_{n}(k_1)\vert^2+E^2)e^{i2\pi n\beta}}{t_2}  & Ee^{2i\pi(2n-1)\beta}\\
-E &-t_2e^{i2\pi(n-1)\beta}  \end{pmatrix},\ c_{n}(k_1)=1+t_1e^{-i(k_1+2\pi n\beta)},\ t_1\neq 1.
\end{equation}
The determinant of $T_{n}(k_1)$ is
\begin{equation}
    \det T_{n}(k_1)=\bar{c}_{n}(k_1)/c_{n}(k_1)\cdot e^{i2\pi(2n-1)\beta},
\end{equation}
which satisfies $|\det T_{n}(k_1)|=1$, but $\det T_{n}(k_1)\neq 1$.
From the definition of $m$ in Appendix A, we get $m=0$ for the transfer matrix $T=\lim\limits_{n\to \infty} T_n\cdots T_1$.
Thus, for the hexagonal Hofstadter model, $\xi=1/\gamma(T)$, i.e., the localization length is equal to the inverse of LE.

 We now prove that the LE $\gamma(T)\geq 0$ for the hexagonal lattice. We define $T'_n(k_1)\equiv
\frac{T_{n}(k_1)}{\sqrt{\det T_{n}(k_1)}}$, which is then $SL(2,\mathbb{C})$. The LE of $T'$ is then non-negative, i.e., 
$\gamma(T')\geq 0$. From the definition of $T'_n(k_1)$, we can see that $\gamma(T')=\gamma(T)-m$. Since $m=0$ for the transfer matrix $T$ of the hexagonal lattice, we get $\gamma(T)\geq 0$.

From Eq.(\ref{gamma}) in the main text, 
\begin{equation}
    \gamma(T)=\gamma(B)-I.
\end{equation}
We then get $\gamma(B)\geq I$ at $\epsilon\to 0$.

\section{Calculation of the Integral $I^\epsilon= \frac{1}{2\pi}\int_0^{2\pi}  \ln\vert c_{j}(k_1+2\pi i\epsilon)\vert d k_1 $}

In this appendix, we calculate the integal $I^\epsilon$ in Eq.(\ref{Jensen}), i.e.,  
\begin{equation}
I^\epsilon=\frac{1}{2\pi}\int_0^{2\pi} \ln\big|c_j(k_1+2\pi i\epsilon)\big| dk_1
\end{equation}
with the complex function
$c_j(k_1+2\pi i\epsilon)
= 1 + t_1 e^{-i(k_1+2\pi n\beta+2\pi i\epsilon)},$
where \(t_1>0\), and \(n,\beta,\epsilon\in\mathbb{R}\) are real constants. 

We denote $\theta\equiv k_1+2\pi n\beta$,  
and use Jensen's formula  for a function \(f(z)\) analytic on the closed unit disk \(|z|\le 1\):
\begin{equation}
\frac{1}{2\pi}\int_0^{2\pi} \ln\big|f(e^{i\theta})\big|d\theta
= \ln|f(0)| + \sum_{a_k} \ln\frac{1}{|a_k|},
\end{equation}
where \(\{a_k\}\) are the zeros of \(f(z)\) inside the unit disk \(|z|<1\). 
Define the analytic function $
f(z) = 1 + t_1 e^{2\pi\epsilon}\,z.$
The only zero of $f(z)$ is at
$
z_0 = -\frac{1}{t_1 e^{2\pi\epsilon}}.
$
We distinguish three cases:

\textbf{Case 1}: \(t_1 e^{2\pi\epsilon} < 1\).
The zero \(z_0\) lies outside the unit disk (\(|z_0|>1\)), so there are no zeros inside \(|z|<1\).
Jensen's formula gives
\begin{equation}
\frac{1}{2\pi}\int_0^{2\pi} \ln\big|c_j(k_1+2\pi i\epsilon)\big| dk_1= \ln f(0) = 0.
\end{equation}

\textbf{Case 2}: \(t_1 e^{2\pi\epsilon} > 1\).
The zero \(z_0\) lies inside the unit disk (\(|z_0|<1\)). Using \(f(0)=1\) and Jensen's formula, we obtain
\begin{equation}
\frac{1}{2\pi}\int_0^{2\pi} \ln\big|c_j(k_1+2\pi i\epsilon)\big| dk_1 = \ln\big(t_1 e^{2\pi\epsilon}\big) = \ln t_1 + 2\pi\epsilon.
\end{equation}

\textbf{Case 3:} $t_1e^{2\pi \epsilon}=1.$ The zero $z_0$ lies on the unit disk ($|z_0|=1$). Using $\int_{0}^{\frac{\pi}{2}}\ln (\cos \phi)\ d\phi=-\frac{\pi}{2}\ln 2$, we obtain

\begin{align}
    \frac{1}{2\pi}\int_{0}^{2\pi} \ln \vert c_{j}(k_1+2\pi i\epsilon_{0})\vert dk_1&= \ln 2+\frac{1}{2\pi}\int_{0}^{2\pi}\ln \vert \cos (\frac{k_1+2\pi n\beta}{2})\vert dk_1\notag\\ 
    &=\ln 2 +\frac{1}{\pi}\int_{0}^{\pi} \ln \vert \cos \phi\vert\  d\phi\notag\\ 
    &=0.
\end{align}

Combining three cases, the final result reads
\begin{equation} 
I^\epsilon=\frac{1}{2\pi}\int_0^{2\pi} \ln\big|c_j(k_1+2\pi i\epsilon)\big| dk_1 =
\begin{cases}
0, & t_1 e^{2\pi\epsilon} \leq 1, \\[4pt]
\ln t_1 + 2\pi\epsilon, & t_1 e^{2\pi\epsilon} > 1,
\end{cases}
\end{equation}
and $I^\epsilon \geq 0$.

\end{widetext}

\newpage


\begin{thebibliography}{99}
\bibitem{Hofstadter_1976} D. R. Hofstadter, \textit{Energy levels and wave functions of Bloch electrons in rational and irrational magnetic fields}, Phys. Rev. B {\bf 14}, 2239 (1976).

\bibitem{Von-Klitching} K. v. Klitzing, G. Dorda, and M. Pepper, \textit{New Method for High-Accuracy Determination of the Fine-Structure Constant Based on Quantized Hall Resistance}, Phys. Rev. Lett. {\bf 45}, 494 (1980).

\bibitem{Hofstadter_k_Bernevig_2020} J. Herzog-Arbeitman, Z.-D. Song, N. Regnault, and B. A. Bernevig, \textit{Hofstadter Topology: Noncrystalline Topological Materials at High Flux}, Phys. Rev. Lett. {\bf 125}, 236804 (2020).

\bibitem{Hexagonal_Wangjian_2020} J. Wang and L. H. Santos, \textit{Classification of Topological Phase Transitions and van Hove Singularity Steering Mechanism in Graphene Superlattices}, Phys. Rev. Lett. {\bf 125}, 236805 (2020).

\bibitem{WEN1989} X. G. Wen and A. Zee, \textit{Winding number, family index theorem, and electron hopping in a magnetic field}, Nucl. Phys. B {\bf 316}, 641 (1989).

\bibitem{Xiao2024} R. Xiao and Y. X. Zhao, \textit{Revealing the spatial nature of sublattice symmetry}, Nat. Commun. {\bf 15}, 3787 (2024).

\bibitem{Gao_Chen_2026} Q. Gao and W. Chen, \textit{Topological constraints on the electronic band structure of a hexagonal lattice in a magnetic field}, Phys. Rev. B {\bf 113}, 205118 (2026).

\bibitem{Square(NNN)_Kohmoto_1990} Y. Hatsugai and M. Kohmoto, \textit{Energy spectrum and the quantum Hall effect on the square lattice with next-nearest-neighbor hopping}, Phys. Rev. B {\bf 42}, 8282 (1990).

\bibitem{Zeromodes_Kohmoto_1989} M. Kohmoto, \textit{Zero modes and the quantized Hall conductance of the two-dimensional lattice in a magnetic field}, Phys. Rev. B {\bf 39}, 11943 (1989).

\bibitem{Symmetry_indicators} Y. Fang and J. Cano, \textit{Symmetry indicators in commensurate magnetic flux}, Phys. Rev. B {\bf 107}, 245108 (2023).

\bibitem{Dimerized} A. Lau, C. Ortix, and J. van den Brink, \textit{Topological Edge States with Zero Hall Conductivity in a Dimerized Hofstadter Model}, Phys. Rev. Lett. {\bf 115}, 216805 (2015).

\bibitem{Hofstadter_Real} J. Herzog-Arbeitman, Z.-D. Song, L. Elcoro, and B. A. Bernevig, \textit{Hofstadter Topology with Real Space Invariants and Reentrant Projective Symmetries}, Phys. Rev. Lett. {\bf 130}, 236601 (2023).

\bibitem{Zak} J. Zak, \textit{Magnetic Translation Group}, Phys. Rev. {\bf 134}, A1602 (1964).

\bibitem{Many_Chern} A. Matsugatani, Y. Ishiguro, K. Shiozaki, and H. Watanabe, \textit{Universal Relation among the Many-Body Chern Number, Rotation Symmetry, and Filling}, Phys. Rev. Lett. {\bf 120}, 096601 (2018).

\bibitem{Thoulss_Bandwidths_1983} D. J. Thouless, \textit{Bandwidths for a quasiperiodic tight-binding model}, Phys. Rev. B {\bf 28}, 4272 (1983).

\bibitem{Thouless1990} D. J. Thouless, \textit{Scaling for the discrete Mathieu equation}, Commun. Math. Phys. {\bf 127}, 187 (1990).

\bibitem{thouless1991total} D. J. Thouless and Y. Tan, \textit{Total bandwidth for the Harper equation. III. Corrections to scaling}, J. Phys. A: Math. Gen. {\bf 24}, 4055 (1991).

\bibitem{APT_Thouless_1994} J. H. Han, D. J. Thouless, H. Hiramoto, and M. Kohmoto, \textit{Critical and bicritical properties of Harper's equation with next-nearest-neighbor coupling}, Phys. Rev. B {\bf 50}, 11365 (1994).

\bibitem{S.J_PRL_1995} R. del Rio, S. Jitomirskaya, Y. Last, and B. Simon, \textit{What is Localization?}, Phys. Rev. Lett. {\bf 75}, 117 (1995).

\bibitem{S.J_PRL_1996} S. Ya. Jitomirskaya and Y. Last, \textit{Dimensional Hausdorff Properties of Singular Continuous Spectra}, Phys. Rev. Lett. {\bf 76}, 1765 (1996).

\bibitem{AAH_1980} S. Aubry and G. André, \textit{Analyticity breaking and Anderson localization in incommensurate lattices}, Proc. Int. Colloq. Group-Theor. Methods Phys. {\bf 3} (1980).

\bibitem{Unification} X.-C. Zhou, B.-C. Yao, Y. Wang, Y. Wang, Y. Wei, Q. Zhou, and X.-J. Liu, \textit{The fundamental localization phases in quasiperiodic systems: a unified framework and exact results}, Sci. Bull. {\bf 71}, 1654 (2026).

\bibitem{Jitomirskaya2012} S. Jitomirskaya and C. A. Marx, \textit{Analytic Quasi-Perodic Cocycles with Singularities and the Lyapunov Exponent of Extended Harper's Model}, Commun. Math. Phys. {\bf 316}, 237 (2012).

\bibitem{SquareNNN_2017} A. Avila, S. Jitomirskaya, and C. A. Marx, \textit{Spectral theory of extended Harper's model and a question by Erdős and Szekeres}, Invent. Math. {\bf 210}, 283 (2017).

\bibitem{Avila2015} A. Avila, \textit{Global theory of one-frequency Schrödinger operators}, Acta Math. {\bf 215}, 1 (2015).

\bibitem{Simon_2021} B. Simon, R. Han, S. Jitomirskaya, and M. Zworski, \textit{Honeycomb structures in magnetic fields}, J. Phys. A: Math. Theor. {\bf 54}, 345203 (2021).

\bibitem{B.Hetenyi_1} B. Het\'enyi and I. Balogh, \textit{Numerical study of the localization transition of Aubry-Andr\'e type models}, Phys. Rev. B {\bf 112}, 144203 (2025).

\bibitem{B.Hetenyi_2} B. Het\'enyi, \textit{Scaling of the bulk polarization in extended and localized phases of a quasiperiodic model}, Phys. Rev. B {\bf 110}, 125124 (2024).

\bibitem{Quasicrystal} Y. E. Kraus and O. Zilberberg, \textit{Topological Equivalence between the Fibonacci Quasicrystal and the Harper Model}, Phys. Rev. Lett. {\bf 109}, 116404 (2012).

\bibitem{RG} M. Gonçalves, B. Amorim, E. V. Castro, and P. Ribeiro, \textit{Renormalization group theory of one-dimensional quasiperiodic lattice models with commensurate approximants}, Phys. Rev. B {\bf 108}, L100201 (2023).

\bibitem{RG2} M. Gonçalves, B. Amorim, E. V. Castro, and P. Ribeiro, \textit{Critical Phase Dualities in 1D Exactly Solvable Quasiperiodic Models}, Phys. Rev. Lett. {\bf 131}, 186303 (2023).

\bibitem{Moire_Lattice} C. R. Dean, L. Wang, P. Maher, C. Forsythe, F. Ghahari, Y. Gao, J. Katoch, M. Ishigami, P. Moon, M. Koshino, T. Taniguchi, K. Watanabe, K. L. Shepard, J. Hone, and P. Kim, \textit{Hofstadter's butterfly and the fractal quantum Hall effect in moiré superlattices}, Nature {\bf 497}, 598 (2013).

\bibitem{graphene_superlattice} L. A. Ponomarenko, R. V. Gorbachev, G. L. Yu, D. C. Elias, R. Jalil, A. A. Patel, A. Mishchenko, A. S. Mayorov, C. R. Woods, J. R. Wallbank, M. Mucha-Kruczynski, B. A. Piot, M. Potemski, I. V. Grigorieva, K. S. Novoselov, F. Guinea, V. I. Fal'ko, and A. K. Geim, \textit{Cloning of Dirac fermions in graphene superlattices}, Nature {\bf 497}, 594 (2013).

\bibitem{Heterorstructures} B. Hunt, J. D. Sanchez-Yamagishi, A. F. Young, M. Yankowitz, B. J. LeRoy, K. Watanabe, T. Taniguchi, P. Moon, M. Koshino, P. Jarillo-Herrero, and R. C. Ashoori, \textit{Massive Dirac Fermions and Hofstadter Butterfly in a van der Waals Heterostructure}, Science {\bf 340}, 1427 (2013).

\bibitem{VHS} J. Wang and L. H. Santos, \textit{Classification of Topological Phase Transitions and van Hove Singularity Steering Mechanism in Graphene Superlattices}, Phys. Rev. Lett. {\bf 125}, 236805 (2020).

\bibitem{Note} D. Damanik, \textit{The Spectrum of the Almost Mathieu Operator}, arXiv:0908.1093 (2009).

\bibitem{Anomalous_mobility_edges} T. Liu, X. Xia, S. Longhi, and L. Sanchez-Palencia, \textit{Anomalous mobility edges in one-dimensional quasiperiodic models}, SciPost Phys. {\bf 12}, 027 (2022).

\bibitem{MosaicModel_2020} Y. Wang, X. Xia, L. Zhang, H. Yao, S. Chen, J. You, Q. Zhou, and X.-J. Liu, \textit{One-Dimensional Quasiperiodic Mosaic Lattice with Exact Mobility Edges}, Phys. Rev. Lett. {\bf 125}, 196604 (2020).

\bibitem{Critical_Phases} X.-C. Zhou, Y. Wang, T.-F. J. Poon, Q. Zhou, and X.-J. Liu, \textit{Exact New Mobility Edges between Critical and Localized States}, Phys. Rev. Lett. {\bf 131}, 176401 (2023).



\bibitem{MBL_Numberical} S. Iyer, V. Oganesyan, G. Refael, and D. A. Huse, \textit{Many-body localization in a quasiperiodic system}, Phys. Rev. B {\bf 87}, 134202 (2013).

\bibitem{MBL_Experiment} M. Schreiber, S. S. Hodgman, P. Bordia, H. P. Lüschen, M. H. Fischer, R. Vosk, E. Altman, U. Schneider, and I. Bloch, \textit{Observation of many-body localization of interacting fermions in a quasirandom optical lattice}, Science {\bf 349}, 842 (2015).

\bibitem{MBL_Critical} T. Cookmeyer, J. Motruk, and J. E. Moore, \textit{Critical properties of the ground-state localization-delocalization transition in the many-particle Aubry-Andr\'e model}, Phys. Rev. B {\bf 101}, 174203 (2020).

\bibitem{Experiment1} M. Aidelsburger, M. Atala, M. Lohse, J. T. Barreiro, B. Paredes, and I. Bloch, \textit{Realization of the Hofstadter Hamiltonian with Ultracold Atoms in Optical Lattices}, Phys. Rev. Lett. {\bf 111}, 185301 (2013).

\bibitem{Experiment2} H. Miyake, G. A. Siviloglou, C. J. Kennedy, W. C. Burton, and W. Ketterle, \textit{Realizing the Harper Hamiltonian with Laser-Assisted Tunneling in Optical Lattices}, Phys. Rev. Lett. {\bf 111}, 185302 (2013).

\bibitem{Experiment3} Y. E. Kraus, Y. Lahini, Z. Ringel, M. Verbin, and O. Zilberberg, \textit{Topological States and Adiabatic Pumping in Quasicrystals}, Phys. Rev. Lett. {\bf 109}, 106402 (2012).

\bibitem{Experiment4} Y. Lahini, R. Pugatch, F. Pozzi, M. Sorel, R. Morandotti, N. Davidson, and Y. Silberberg, \textit{Observation of a Localization Transition in Quasiperiodic Photonic Lattices}, Phys. Rev. Lett. {\bf 103}, 013901 (2009).

\bibitem{Data} The plots of fractal dimension analysis in Fig.\ref{FD_analysis} in this manuscript can be produced by the code at  https://doi.org/10.5281/zenodo.19756007 .
\end{thebibliography}
\end{document}